\documentclass[10pt, conference, letterpaper]{IEEEtran}
\usepackage{cite}
\usepackage{amsmath,amssymb,amsfonts}
\usepackage{algorithmic}
\usepackage{graphicx}
\usepackage{xcolor}
\usepackage{amsthm}

\newtheorem{theorem}{Theorem}
\newtheorem{lemma}{Lemma}
\newtheorem{corollary}{Corollary}

\def\BibTeX{{\rm B\kern-.05em{\sc i\kern-.025em b}\kern-.08em
    T\kern-.1667em\lower.7ex\hbox{E}\kern-.125emX}}
\begin{document}

\title{
Optimal Allocation of Tasks and Price of Anarchy of Distributed Optimization\\
in Networked Computing Facilities
}

\author{
\IEEEauthorblockN{Vincenzo Mancuso}
\IEEEauthorblockA{\textit{IMDEA Networks Institute, Madrid, Spain} \\
vincenzo.mancuso@imdea.org}
ORCID: 0000-0002-4661-381X
\and
\IEEEauthorblockN{Paolo Castagno}
\IEEEauthorblockA{\textit{University of Turin, Turin, Italy} \\
paolo.castagno@unito.it}
ORCID: 0000-0002-1349-1844
\and
\IEEEauthorblockN{Leonardo Badia}
\IEEEauthorblockA{\textit{University of Padova, Padua, Italy} \\
leonardo.badia@unipd.it}
ORCID: 0000-0001-5770-1199
\\
\and
\IEEEauthorblockN{\hspace{3cm}Matteo Sereno}
\IEEEauthorblockA{\hspace{3cm}\textit{University of Turin, Turin, Italy} \\
\hspace{3cm}matteo.sereno@unito.it}
\hspace{3cm}ORCID: 0000-0002-5339-3456
\and
\IEEEauthorblockN{Marco Ajmone Marsan}
\IEEEauthorblockA{\textit{IMDEA Networks Institute, Madrid, Spain} \\
marco.ajmone@imdea.org}
ORCID: 0000-0002-9560-7053
}

\maketitle

\begin{abstract}
The allocation of computing tasks for networked distributed services poses a question to service providers on whether centralized allocation management be worth its cost. Existing analytical models were conceived for users accessing computing resources with practically indistinguishable (hence irrelevant for the allocation decision) delays, which is typical of services located in the same distant data center. However, with the rise of the edge-cloud continuum, a simple analysis of the sojourn time that computing tasks observe at the server misses the impact of diverse latency values imposed by server locations. We therefore study the optimization of computing task allocation with a new model that considers both distance of servers and sojourn time in servers. We derive exact algorithms to optimize the system and we show, through numerical analysis and real experiments, that differences in server location in the edge-cloud continuum cannot be neglected. By means of algorithmic game theory, we study the price of anarchy of a distributed implementation of the computing task allocation problem and unveil important practical properties such as the fact that the price of anarchy tends to be small---except when the system is overloaded---and its maximum can be computed with low complexity.
\end{abstract}


\section{Introduction}

Meeting the requirements in terms of latency experienced by individual network users is of fundamental importance to achieve the desired quality of service for many real-time applications \cite{zhang2019mobile,wang2023road}. 
A paradigm often adopted in current mobile Internet architectures to tackle this issue is that of edge computing, i.e., bringing computing resources closer to end users, rather than processing data in a centralized location like the cloud \cite{luo2021resource}.
However, as the mobile Internet becomes more and more pervasive, the management of distributed computing and network infrastructure is evolving towards an edge-cloud continuum, rather than a simple dychotomy between edge or cloud computing \cite{cheng2021intelligent}.

Indeed, network users are expected to more and more frequently access services ubiquitously and persistently, through a vast and diverse collection of interconnected autonomous systems, traversing local networks and global backbones \cite{szymanski2014ultra}. Network segments may cover different distances and have their own unique characteristics in terms of transmission technologies, not to mention local management strategies that lead to an overall different handling of traffic.
All of this results in heterogeneous latency expressions for data traffic \cite{RanjanKK04}.

From the standpoint of an individual user, the problem is limited to the choice of the best (i.e., minimum latency) path. When a global perspective is adopted, establishing coordination among multiple users becomes of formidable complexity and is practically infeasible. The crux of the matter becomes whether distributed approaches to server selection in extremely variegate network architectures can still be efficient \cite{BELL_STIDHAM}.

While this problem received attention in the past \cite{Qiu2003,Fraleigh2003,Sen2004},
we argue that the available mathematical frameworks are inadequate to represent the edge-cloud continuum, for the following reasons.
First of all, most of the investigations consider some model for the latency that is kept homogeneous among the alternatives \cite{HAVIV_ROUGH}. Despite accounting for possible differences among multiple service options, the comparisons involve, e.g., fast vs slow servers but just with different parameters in the same formula. Instead, the service alternatives in the mobile Internet have characteristics that are different not just in quantitative but also in qualitative terms.

One particular instance of this aspect is the fixed component of the latency, which comes for the most part from the physical distance of the server location. We argue that the edge-cloud continuum is exploiting servers at extremely different locations on a global scale \cite{milojicic2020edge}, and therefore there may be fixed latencies to reach the server that are actually independent of the congestion at the server itself, or how fast the server is. We will show how neglecting this aspect leads to extremely suboptimal choices.

To complicate things further, one can observe that, while the absolute service capacity in the absence of congestion and/or the fixed latencies due to distance can be possibly known to the user, the final performance also depends on the congestion at the chosen server, which is harder to estimate \cite{wang2012congestion}. 

The direction that we explore in this paper is that of algorithmic game theory \cite{Roughgarden2002}, considering the service choices performed by a multitude of atomic selfish non-cooperative network users interested in minimizing their own latency.
In particular, we will derive the price of anarchy (PoA) \cite{KoutsoupiasPapadimitriou2009} of a distributed allocation of computing tasks in heterogeneous scenarios. Compared to existing results available in the literature, we take a general approach that can be applicable to any functional relationship describing the achieved latency, under very mild assumptions of positive first and second derivative.
We verify our analytical approach by verifying that it obtains known results in special cases, but at the same time we argue that the location of servers in the network must be properly taken into account to obtain satisfactory performance. Also, we validate our quantitative results with real-world experiments, which we are able to match thanks to the generality of our approach.

In more detail, this paper presents multiple contributions as follows. First, we evaluate the efficiency of distributed choices by network users in the edge-cloud continuum, under an extremely general framework that is not found in the previous literature, where only special cases have been considered. In particular, we investigate the Nash equilibrium (NE) of the distributed selections, which is found to be unique. 
From this perspective, we assert that this paper can be regarded as a valuable contribution akin to those emphasized in \cite{GHOSH20211}.

Second, we formulate exact algorithmic solutions to derive the optimal allocation point and the Nash equilibrium. Third, we compute the PoA as a function of the network load. The worst-case PoA is proven to be the maximum among a finite number of cases, happening at the transitions between server activation points. Finally, we perform an extensive numerical evaluation, which is also corroborated by experimental results.

The rest of this paper is organized as follows.
In Section \ref{RELATED}, we discuss the related literature. The analysis of the optimal allocation, NEs, and resulting PoA is presented in Section \ref{GENERAL} under very general assumptions. In Section \ref{SPECIAL}, we prove that our analysis can, among other things, generalize the results obtained in the literature for some special cases. We present numerical evaluations in Section \ref{RESULTS} and we finally conclude the paper in Section \ref{CONCS}.

\section{Related Work\label{RELATED}}
Our version of the server selection problem considers an infinite flow of arrivals following a Poisson process. 
The arriving customers are required to choose one of the available servers in the system, each represented by a queueing system with a different service rate. Additionally, a fixed delay represents 
the time it takes for a customer to reach the selected server.

These characterizations find a match in the literature at many levels. The problem as a whole can be seen as a specific instance of traffic routing models \cite{Sheffi1985}. 
The research in transportation optimization, pioneered by Pigou \cite{Pigou1920} and Braess \cite{Braess1969},  achieved numerous important results that also apply to computer networks, scheduling, routing, and optimization of server selection (see \cite{Roughgarden2005} or \cite{Feldmann2003} for detailed discussions).

Regarding the server selection problem, our study is related to the proposals of Bell and Stidham \cite{BELL_STIDHAM}, 
Haviv and Roughgarden \cite{HAVIV_ROUGH}, 
and Wu and Starobinski \cite{UNO_STAROBISKY, WuStarobinski2008}. 
All of these approaches aim to optimize the customer mean waiting time and compare the selfish strategy, 
where customers choose servers to minimize their individual waiting times, with the social (i.e., global) optimization of the average customer mean waiting time. \textcolor{black}{Here is where we provide novel analytical and algorithmic results by casting the original server selection problem over the edge-cloud continuum scenario.}

In~\cite{BELL_STIDHAM}, 
the authors describe the structure of the solutions for individual and social optimizations, 
comparing the two strategies. The study assumes each server is characterized by two parameters: the mean and the second moment 
of the service time distribution. Notably, \textcolor{black}{and differently from what we derive in this work,} most of the results presented in~\cite{BELL_STIDHAM} assume all servers share a common coefficient of variation, facilitating the mathematical derivations.

The other two approaches mentioned earlier use the PoA \cite{KoutsoupiasPapadimitriou2009}
to measure the inefficiency between selfish and social optimization. 
In both cases, closed forms are derived for the two types of optimization, 
as they assume that the service times of different servers are characterized by negative exponential distributions. \textcolor{black}{We show that in presence of more complex and heterogeneous networks, closed forms are still possible to find, but only for simplified cases. Moreover, we show that an exact algorithmic approach is possible for more general cases and exhibits low-order polynomial complexity.}

In summary, the three mentioned approaches suggest that when each server selection choice 
can be characterized by a single parameter (e.g., average server service time), 
analytical expressions for the two optimizations can be derived and the PoA can be studied. 
In other words, to maintain analytical tractability, they assume that the end-to-end delay 
experienced by service requests is dominated by server delays, but we show that this assumption is not reasonable in the cloud-continuum and requires a different analysis. 

Some other studies support the assumption that the delay experienced by customers is practically due to server processing only. For instance,~\cite{Qiu2003, RanjanKK04, Fraleigh2003, Sen2004} consider the case of server selection on the Internet, where a fundamental role is played by the 
network over-provisioning that generates scenarios where network delays are homogeneous.
However, it is essential to point out that the edge-cloud paradigm is becoming pervasive on a global scale, causing extremely near as well as very far servers to coexist as available option. Together with the increase of processing capabilities, this makes the network delay a significant performance bottleneck \cite{szymanski2014ultra}. Thus, we argue that neglecting this aspect in server selection is
potentially prone to significant errors.

This forms our primary motivation: to compare individual (selfish) and social optimization when network 
delays cannot be neglected. In our study, each path (e.g., server selection) is characterized 
by \textit{at least} two parameters: average server service time and network delay required to reach that server.

We would like to emphasize that extensive research has been conducted on the server selection problem, yielding recent findings. 
Notably, investigations have been carried out on systems employing processor sharing policies in the queueing systems \cite{Altman2011}. 
Server selection strategies assuming knowledge of the queueing process and utilizing the join-the-shortest-queue policy have received 
considerable attention in numerous studies \cite{Foley2001,GUPTA2007}. 

Additionally, a comprehensive recent survey paper \cite{GHOSH20211} 
provides a summary of both recent and earlier studies, 
examining the efficiency loss between selfish and optimized strategies that is a typical issue of the server selection problem.
\textcolor{black}{
The results we present in this paper provide valuable insight for a better understanding of the causes of inefficiency of distributed (selfish) solutions with respect to centrally optimized solutions, thus offering potential benefits to the applications of PoA-based analysis (e.g., the studies presented in \cite{Foley2001,GUPTA2007,GHOSH20211}).}

\begin{figure}
    \centering
    \includegraphics[width=0.9\columnwidth]{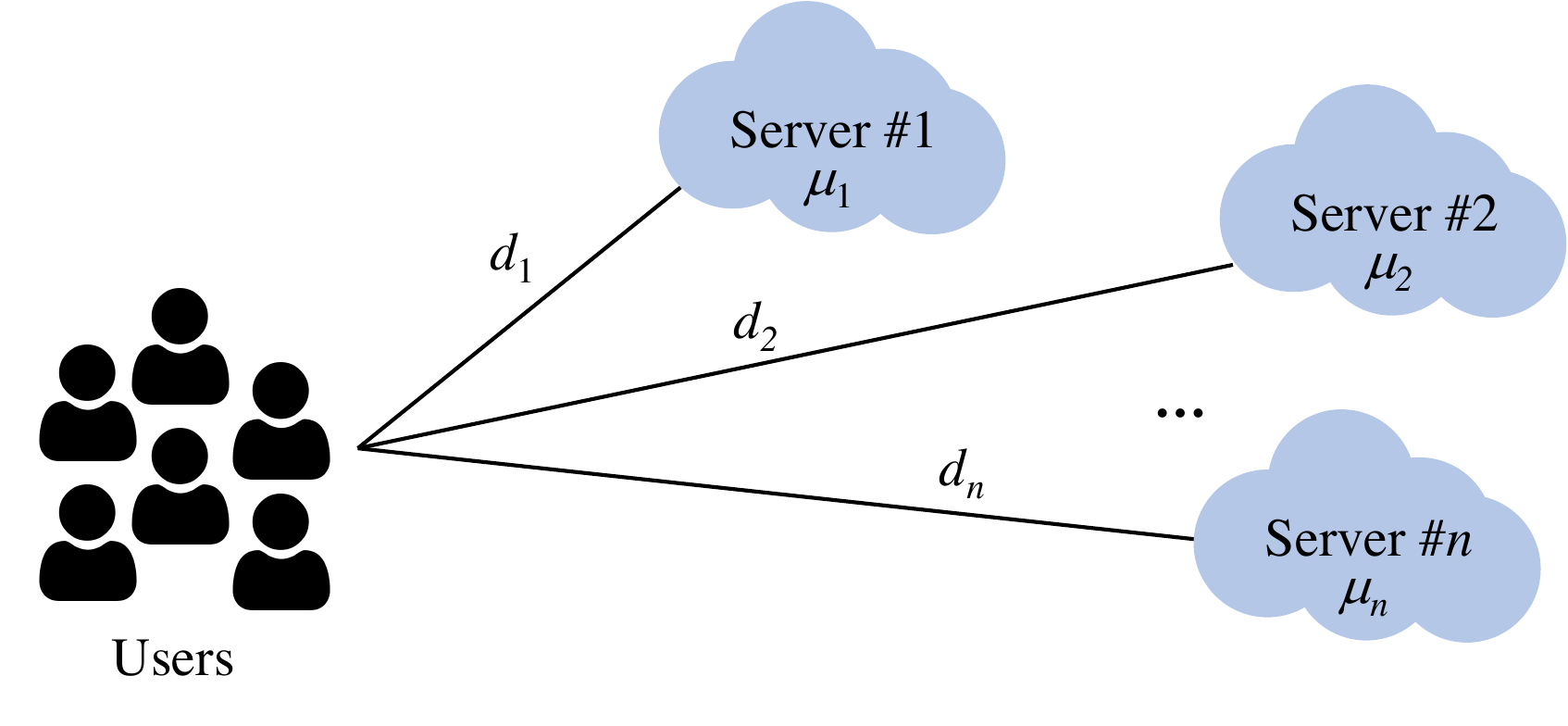}
    \vspace{-2mm}
    \caption{Reference scenario: a group of users on a single network slice with computing resources in the edge-cloud continuum}
    \label{fig:scenario}
\end{figure}

\section{General analysis}
\label{GENERAL}

\textcolor{black}{Assume that a group of users is connected to the same access network (e.g., the same Internet service provider), from which $n$ servers can be reached, as shown in Fig.~\ref{fig:scenario}. Although the servers are accessible at different distances, they belong to the same network slice, which has been created and deployed over the edge-cloud continuum to offer a specific service to the group of users, as typical of modern 3GPP networks~\cite{3gpp_TS23501,3gpp_TS28531}.} The servers are ordered according to the minimum average delay experienced by a job sent to each server, determined by the combined sum of path latency and average service time, excluding any queuing delays, 
\begin{align}
    \forall i,j \in \mathcal{S} = \{1,\cdots,n\}, \; i < j \implies d_i {+} \frac {1}{\mu_i} \le d_j {+} \frac {1}{\mu_j},
    \label{eq:order}
\end{align}
where $d_i$ is the two-way delay to reach server $i$ and $\mu_i$ is the average service time at the same server.

In the following, let $\mathbf{p}^{\star}$ denote a probability vector that expresses the solution of the optimization problem consisting in allocating load to servers such that the average latency be minimized. 
In addition, let $\mathbf{p}^{\dagger}$ denote the Nash equilibrium point (NEP) that would be achieved if users could select their server selfishly, with a stochastic strategy  aiming at minimizing their expected latency. 

\subsection{Optimal load allocation problem}
 
The problem to solve consists in finding the probabilities $\mathbf{p}^{\star} = \{p_i\}_{i=1,\cdots, n}$ that minimize the following cost function:
\begin{align}
    U(\mathbf p) = \sum_{i=1}^n p_i \,\ell_i(p_i\,\Lambda),
    \label{eq:utility_general}
\end{align}
where $\ell_i(x) > 0$ is the average latency of server $i$ when it receives traffic with intensity $0 \le x < \mu_i$, and $\mu_i$ is the capacity of that server. $\Lambda$ is the aggregate offered traffic.

Function $\ell_i(x)$ accounts for the path latency to reach the server ($d_i$), its average service capacity ($\mu_i$), and can also account for additional parameters (e.g., the variance of the service time). The minimum average latency must be the sum of distance plus one service interval, i.e., $\ell_i(0) = d_i + 1/\mu_i$.

Since we consider systems subject to congestion, $\ell_i(x)$ monotonically increases with $x$. Also, $\ell_i(x)$ is a convex functions, as commonly observed for the latency of a queueuing system. Hence, $\ell_i(x)$ and its first and second derivatives are positive functions. 

The above problem is solved subject to the following simple and self-explanatory constraints: 
\begin{align}
    & p_i \ge 0, \quad 
    p_i \Lambda \le \mu_i, \quad \forall i; 
    \quad \sum_i p_i = 1.
    \nonumber 
\end{align}

The problem is strictly convex because $U(\mathbf p)$ has a (diagonal) positive-definite Hessian. Thus, the solution is unique. 

The Lagrangian, with $\alpha_i$, $\beta_i$, and $\gamma$  multipliers, is 
\begin{align}
    \mathcal{L} {=} 
    \sum_i \! p_i \ell_i(p_i \Lambda)
    {-} \! \sum_i \! \alpha_i p_i 
    {+} \! \sum_i \! \beta_i \left(p_i \Lambda {-} \mu_i\right)
    {+} \gamma \!\left(\!1{-}\!\sum_i p_i\!\right)\!. \nonumber 
\end{align}

The following KKT conditions must hold and give necessary conditions for a vector $\mathbf{p^{\star}}$ to be a solution of the problem: 
\begin{align*}
    & \frac{\partial \mathcal{L}}{\partial p_i} 
    = \ell_i(p_i\, \Lambda) {+} p_i \,\Lambda \left. \frac{d\, \ell_i(x)}{d\, x} \right|_{x=p_i\,\Lambda} \!\!\!\!{-} \alpha_i {+} \Lambda \beta_i {-} \gamma = 0, \quad \forall i;
     \\
  & \alpha_i p_i  = 0, \quad \beta_i \left( p_i \Lambda - \mu_i \right) = 0, \quad \forall i; 
\\
    & \gamma \left(1 - \sum_i p_i \right) = 0.
\end{align*}

Notice that $\ell_i(x) + p_i \,\Lambda \frac{d\, \ell_i(x)}{d\, x}$ is the derivative of the weighted latency of the $i$-th server, with weight $p_i$. Multipliers $\alpha_i$ and $\beta_i$ must be non-negative, while $\gamma$ can take any real value (because it is associated to an equality constraint). 

In order to identify which servers are used, consider the properties of the KKT conditions for the servers at the optimal point $\mathbf{p}^{\star}$, which is a solution of the formulated problem. 

\textbf{Case $p_j^{\star} = 0$.}
If the optimal solution consists in assigning zero load to server $j$, then the KTT conditions imply that
\begin{align}
    & \alpha_j \ge 0, \quad \beta_j = 0, \quad \ell_j(0) - \alpha_j - \gamma = 0;
    \\
    & \Rightarrow
    \ell_j(0) \ge \gamma,
    \label{eq:gamma_out_generic}
\end{align}
i.e., servers that do not need to be active are those for which the latency computed at empty queue satisfies the above condition. However, with no queueing, a job is served in $\ell_j(0) = d_j + 1/\mu_j$, for any function $\ell_j$. Therefore, if a server $j$ is not active, then any node $i>j$, for which we know that $d_i + 1/\mu_i \ge d_j + 1/\mu_j$, is also inactive. 

Being the set of servers ordered, there exists an integer $j^{\star} \le n$ such that all servers from 1 to $j^{\star}$ receive some traffic, while the remaining servers remain inactive.

Finding the value of $j^{\star}$ is part of the solution of the optimization problem and we will soon show how to find it. 

Next, analyze the KKT conditions for nodes that have to receive non-zero traffic. 

\textbf{Case $p_j^{\star} > 0$.}
For all nodes $j = 1,\cdots, j^{\star}$, which receive non-zero load in the optimal configuration, the following conditions must hold: 
\begin{align}
    \alpha_j & = \beta_j = 0, \\
    \gamma & = \ell_j(p_j^{\star}\, \Lambda) + p_j^{\star} \, \Lambda \left . \frac{d\, \ell_j(x)}{d\, x}\right | _ {x = p_j ^{\star} \, \Lambda}.
    \label{eq:gamma_in_generic}
\end{align}
Therefore, $\gamma$ must be positive and, since $p_j^{\star}>0$, must also be greater than any $\ell_j(0)$ for any active server, which leads to 
\begin{align}
\gamma > d_{j^{\star}} + \frac{1}{\mu_{j^{\star}}}.
 \label{eq:gamma_min_generic}
\end{align}
The above result is coherent with \eqref{eq:gamma_out_generic}, which defines when a server receives no load. 
The result also proves $\gamma$ to be a monotonically increasing,
hence invertible, function $h_j(x)$, to be computed in $x = p_j^{\star}\, \Lambda$ at the optimum. Thus, we have
\begin{align}
p_j^{\star} = \frac{1}{\Lambda} \, h_j^{-1}(\gamma).
\label{eq:pi_inversion_generic}
\end{align}
Considering that the latency functions $\ell_j$ and the corresponding derivatives are defined only for offered traffic comprised between 0 and $\mu_j$, the above function $h_j^{-1}(\gamma)$ can only take values in the interval $[1, \mu_j]$ (or $[1, \mu_j)$, if the latency function diverges at $\mu_j$). Hence, $h_j^{-1}/\Lambda$ cannot be larger than 1. Indeed, the sum of all non-zero probabilities must be 1, so: 
\begin{align}
    1 = 
    \sum_{j = 1}^{j^{\star}} p_j^{\star}  
    =  
    \frac{1}{\Lambda} \sum_{j = 1}^{j^{\star}}
    h_j^{-1}(\gamma).
    \label{eq:gamma_invert_generic}
\end{align}
 Since the sum of monotonic increasing functions is also monotonic and increasing, the R.H.S of~\eqref{eq:gamma_invert_generic} is invertible. In addition, the R.H.S. must be a number between 0 and $\sum_{j=1}^{j^{\star}} \mu_j/\Lambda \ge 1$, so that all the offered traffic can be served. Thus, there exists a unique value of $\gamma$ that satisfies~\eqref{eq:gamma_invert_generic}.

Notice that, because of conditions~\eqref{eq:gamma_out_generic} computed on $j^{\star}{+}1$ and \eqref{eq:gamma_min_generic} for $j^{\star}$, $\gamma$ must be found in a well identified interval:
\begin{align}
\gamma \in \left(d_{j^{\star}} + \frac{1}{\mu_{j^{\star}}}, d_{j^{\star}+1} + \frac{1}{\mu_{j^{\star}+1}}\right],
\label{eq:gamma_interval_generic}
\end{align}
where $d_{j^{\star}+1} {+} \frac{1}{\mu_{j^{\star}+1}}$ has to be taken as infinite if $j^{\star} {=} n$. Hence, the number of active servers $j^{\star}$ is uniquely identified by $\gamma$. 

In practice, determining the value of $j^{\star}$ beforehand is key to compute the solution of the problem efficiently. Indeed, once $j^{\star}$ is known, by inverting the normalization expression~\eqref{eq:gamma_invert_generic}, one can compute the value of $\gamma$, and, in turn, the value of $\gamma$ can be used in~\eqref{eq:pi_inversion_generic} to compute the optimal probabilities $p_j^{\star}$.

Probabilities $p_j^{\star}$ depend on $\Lambda$, although so far we have treated $\Lambda$ as a constant. 
If instead we take $\Lambda$ as a variable, we can see that $\gamma$ must be a monotonically increasing function of such a variable, because so are all functions $\ell_j$ and associated derivatives.
Thus, as $\Lambda$ increases, $\gamma$ increases too, so that, as inferred from~\eqref{eq:gamma_interval_generic}, servers are progressively activated following the order defined by~\eqref{eq:order}. 
In particular, if server $j$ is activated at  $\Lambda^{(\mathrm{Opt})}_j$, with that load $p^{\star}_j$ is still 0 and so, according to~\eqref{eq:gamma_in_generic}, $\gamma = \ell_j(0) = d_j + \frac{1}{\mu_j}$. 
Hence, at $\Lambda_j^{(\mathrm{Opt})}$ we have: 
\begin{align}
p^{\star}_i \, \Lambda_j^{(\mathrm{Opt})}
= 
h_i^{-1} \left(d_j + \frac{1}{\mu_j}\right), \quad \forall i \le j,
\end{align}
 and by summing over servers with non-zero probability $p_i^{\star}$, i.e., from 1 to $j-1$, we obtain the traffic threshold: 
\begin{align}
\Lambda_j^{(\mathrm{Opt})}
= 
\sum_{i=1}^{j-1} h_i^{-1} \left(d_j + \frac{1}{\mu_j}\right).
\label{eq:activation_opt_general}
\end{align}
 
A simple comparison between the activation thresholds $\Lambda_j^{(\mathrm{Opt})}$ and the offered traffic $\Lambda$ thus reveals the value of $j^{\star}$. 
Of course, $\Lambda_1^{(\mathrm{Opt})} {=} 0$ since at least the first server has to be active as soon as non-zero traffic is offered to the network.

Considering the analysis above, the optimization of the formulated problem can be done with an algorithm that requires the numerical inversion of a few monotonic functions, which can be done in closed form only in specific cases, whereas in the general case it requires numerical inversion with lightweight algorithms as the dichotomic search algorithm. 
In particular, the following Exact Probability Mapping (\texttt{EPM}) algorithm finds the optimal probability vector $\mathbf{p^{\star}}$ at which the average system latency is minimized. 

\begin{description}
\item[\tt Input:] $\quad\quad$Sorted servers $j \in \{1,\cdots,n\}$, $\Lambda$.
\item[\tt Step 1:] $\quad\quad$Compute activation thresholds 
$\Lambda_j^{(\mathrm{Opt})}$ with~\eqref{eq:activation_opt_general}.
\item[\tt Step 2:] $\quad\quad$Compute $j^{\star}$ by comparing $\Lambda$ to the thresholds.
\item[\tt Step 3:] $\quad\quad$Set $p_j^{\star}{=}0, \forall j {>} j^{\star}$ and compute $\gamma$ inverting \eqref{eq:gamma_invert_generic}. 
\item[\tt Step 4:] $\quad\quad$Compute $p_j^{\star}$, $\forall j\le j^{\star}$, with \eqref{eq:pi_inversion_generic}.
\item[\tt Output:] $\quad\quad$$p_j^{\star}, j \in \{1,\cdots,n\}$.
\end{description}

\begin{theorem}
The \texttt{EPM} algorithm is exact and  polynomial.
\label{th:complexity}
\end{theorem}

\begin{proof}
The algorithm finds the optimum because it reaches a feasible solution $\mathbf{p^{\star}}$ for which all KKT conditions are satisfied. 

The algorithm finds the activation thresholds---which requires computing a number of terms which is quadratic in $n$---and inverts~\eqref{eq:gamma_invert_generic}. The latter consists in finding the zero of a monotonically increasing function with up to $n+1$ invertible terms, which can be done, e.g., with a dichotomic search on each term, with complexity $\mathcal{O}(n \log r)$, where $r$ is the target numerical resolution. The computation complexity of each of the $n$ probabilities, with \eqref{eq:pi_inversion_generic}, is the complexity of one inversion, i.e., $\mathcal{O}(\log r)$.
Therefore, the overall complexity of the \texttt{EPM} algorithm is $\mathcal{O}(n^2 + n\log r)$ and is polynomial for any value of $r$. 
\end{proof}

\subsection{The Nash equilibrium point (NEP)}
We can consider a distributed version for the minimization of~\eqref{eq:utility_general}, in which each user sending traffic minimizes her latency with a probabilistic strategy $\mathbf{p_u^{\dagger}}$. 

With sorted servers, it is easy to see that a user sends traffic to $j$ only if the average latency she observes is above the minimum possible latency at that server, $\ell_j(0)$. 

A NEP exists and is unique due to fact that 
$U(\mathbf p)$ is strictly convex~\cite{BELL_STIDHAM}. 
At the NEP, all users have the same strategy $\mathbf{p_u^{\dagger}} {=} \mathbf{p}^{\dagger}$ characterized by the fact that all servers that receive traffic experience the same average latency, so that no user has any incentive to deviate from strategy $\mathbf{p^{\dagger}}$:
\begin{align}
    \ell_j(p_j^{\dagger} \, \Lambda) = \alpha, \quad  p_j^{\dagger} >0, \quad \forall j \le j^{\dagger}, 
\end{align}
where $\alpha$ is the average latency in the system, which is the same for all loaded servers, and $j^{\dagger}$ is the number of used servers ($p_j^{\dagger} = 0$ for other servers $j > j^{\dagger}$). 

As $\Lambda$ increases, $\alpha$ has to increase as well because, in order to maintain the same latency at all active servers, the incremental arrival rate has to be distributed over all of them and no server can receive less traffic than before the increase. 
Notice also that, for a used server ($p_j^{\dagger} > 0)$ the latency's lower bound is 
\begin{align}
    \alpha > \ell_j(0) = d_j + \frac{1}{\mu_j}, \quad \forall j \le j^{\dagger}. 
\end{align}
Expressing probabilities vs $\alpha$  and normalizing, we obtain: 
\begin{align}
    p_j^{\dagger} & = \frac{1}{\Lambda} \, \ell_j^{-1} (\alpha), \quad \forall j \le j^{\dagger}; 
    \label{eq:pi_inv_alpha}
    \\
    1 & = \frac{1}{\Lambda} \, \sum_{j=1}^{j^{\dagger}} \ell_j^{-1} (\alpha).
    \label{eq:alpha_inv}
\end{align}
Inverting the above expression yields the value of $\alpha>0$, which must be unique since the R.H.S. of~\eqref{eq:alpha_inv} is a monotonic increasing function taking values between 0 (when $\alpha$ is 0) and $(1/\Lambda) \sum_{j=1}^{j^{\dagger}} \Lambda = j^{\dagger}$ (valid for $\alpha \rightarrow \infty$).  
Notice that $\alpha$ must be always comprised in the following interval:
\begin{align}
    \alpha \in 
        \left(
            d_{j^{\dagger}} + \frac{1}{\mu_{j^{\dagger}}},
            d_{j^{\dagger}+1} + \frac{1}{\mu_{j^{\dagger}+1}}
        \right).
\end{align}
This implies that $\alpha$ has to increase also when a new server is activated because of an increase of $\Lambda$. We conclude that $\alpha$ always increases with $\Lambda$ and servers are progressively activated following the sorting order~\eqref{eq:order}, as $\Lambda$ increases.

Since at activation of $j$ the value of $p_j^{\dagger}$ is 0 and therefore $\alpha = d_j {+} \frac{1}{\mu_j}$, and the sum of non-zero probabilities must be equal to 1, the threshold $\Lambda_j^{(\mathrm{NEP})}$ can be computed similarly to the threshold in the optimal case (of course, $\Lambda_1^{(\mathrm{NEP})}{=}0$): 
\begin{align}
\Lambda_j^{(\mathrm{NEP})} = 
\sum_{i=1}^{j-1} 
\ell_j^{-1}\left(d_i+\frac{1}{\mu_i}\right).
\label{eq:activation_nep_general}
\end{align}

In the NEP calculation, function $\ell_j$ plays the role that $h_j$ plays in the calculation of the optimum. Since both functions are positive and monotonic increasing, the algorithm for the NEP is very similar to the one for the optimum: 

\begin{description}
\item[\tt Input:] $\quad\quad$Sorted servers $j \in \{1,\cdots,n\}$, $\Lambda$.
\item[\tt Step 1:] $\quad\quad$Compute thresholds 
$\Lambda_j^{(\mathrm{NEP})}$ with~\eqref{eq:activation_nep_general}.
\item[\tt Step 2:] $\quad\quad$Compute $j^{\dagger}$ by comparing $\Lambda$ to the thresholds.
\item[\tt Step 3:] $\quad\quad$Set $p_j^{\dagger}{=}0, \forall j {>} j^{\dagger}$ and compute $\alpha$ inverting \eqref{eq:alpha_inv}. 
\item[\tt Step 4:] $\quad\quad$Compute $p_j^{\dagger}$, $\forall j \le j^{\star}$, with \eqref{eq:pi_inv_alpha}.
\item[\tt Output:] $\quad\quad$$p_j^{\dagger}, j \in \{1,\cdots,n\}$.
\end{description}

\begin{theorem}
The NEP calculation algorithm is polynomial.
\end{theorem}
\begin{proof}
The proof proceeds as for Theorem~\ref{th:complexity} and finds complexity $\mathcal{O}(n^2 {+} n\log r)$ for any inversion precision $r$.
\end{proof}

\subsection{Properties}

\begin{theorem}
    $\gamma$ is an upper bound for the optimal latency. 
\end{theorem}
\begin{proof}
    $  
    \!\gamma {=} 
    \sum_{j=1}^{j^{\star}} \!p_j^{\star} \gamma 
    {=}\sum_{j=1}^{j^{\star}} 
    \!p_j^{\star} 
    \!\left(\!
    \ell_j(p_j^{\star} \Lambda) {+} p_j^{\star}   \Lambda \! \left . \frac{d \ell_j(x)}{d x}\right | _ {p_j {=} p_j^{\star} \Lambda }
    \!\right)$ 
    
    $\ge
    \sum_{j=1}^{j^{\star}} p_j^{\star} \ell_j(p_j^{\star} \, \Lambda) = 
    U(\mathbf p^{\star}).
    $
\end{proof}

\begin{lemma}
$\Lambda >0 \implies \gamma > \alpha$.
\label{lemma:ga}
\end{lemma}
\begin{proof}
If $\Lambda{>}0$ we have the NEP at $\mathbf p^{\dagger}$, with average latency $\alpha$, and the optimum (i.e., the minimum latency) at $\mathbf p^{\star}$, with average latency no larger than $\alpha$.  
If $\mathbf p^{\dagger} {\neq} \mathbf p^{\star}$, at the optimum at least one server receives more traffic than at the NEP. Now, since at $\mathbf p^{\dagger}$ all latency values of used servers $\ell_j(p_j^{\dagger} \, \Lambda)$ must be equal to $\alpha$, the latency of servers that are assigned a higher probability at the optimum than at the NEP must be higher than $\alpha$. At the optimum, the server with the highest latency is therefore a server that received more traffic than at the NEP (this holds also for servers that were not active at the NEP), i.e., if
$j^{\prime} = \arg\max_j \ell_j(p_j^{\star} \, \Lambda)$ denotes the server with the highest latency at the optimum, then 
$p_{j^{\prime}}^{\star} {>} p_{j^{\prime}}^{\dagger}$ and hence 
$\ell_{j^{\prime}}(p_{j^{\prime}}^{\star} \Lambda) {>} \ell_{j^{\prime}}(p_{j^{\prime}}^{\dagger} \Lambda) {=} \alpha$. 
However, $\gamma {>}  \ell_j(p_j^{\star} \Lambda)$ for all active servers, hence $\gamma$ is greater than the largest latency of the used servers, which in turn must be larger than $\alpha$. 

The proof completes by considering the remaining case $\mathbf p^{\star} {=}\mathbf p^{\dagger}$, for which  
$\alpha {=} \ell_j(p_j^{\dagger} \Lambda) {=} \ell_j(p_j^{\star} \Lambda)$ at any active server. However, $\gamma {>} \ell_j(p_j^{\star} \Lambda)$ at any $j$ for which $p_j^{\star} {>} 0$, then $\gamma {>} \alpha$. 
\end{proof}

\begin{theorem}
    $\Lambda_j^{(\mathrm{NEP})} \ge \Lambda_j^{(\mathrm{Opt})}, \forall j \in \{1, \cdots, n\}$.
\end{theorem}

\begin{proof}

For $j=1$, both thresholds at the optimum and at the NEP are 0. If there are multiple servers with the same value of $d_j + 1/\mu_j = d_1 + 1/\mu_1$, all of them have to comply with the same activation conditions, so that they are all activated as soon as non-zero traffic is offered, together with $j=1$. Hence, their thresholds are all 0 both at the optimum and in the NEP configuration, and the theorem holds for this subset. 

For all other servers with $d_j + 1/\mu_j > d_1 + 1/\mu_1$, which must be activated for some positive values of the thresholds, $\gamma|_{\Lambda {=} \Lambda_j^{(\mathrm{Opt})}} {=} \ell_j(0)$ 
and $\alpha|_{\Lambda {=} \Lambda_j^{(\mathrm{NEP})}} {=} \ell_j(0)$, i.e., they are equal. 
However, both $\alpha$ and $\gamma$ increase with $\Lambda$ and $\gamma {>} \alpha$ for  $\Lambda>0$ (Lemma~\ref{lemma:ga}), 
so that the above equality can only hold for $\Lambda_j^{(\mathrm{NEP})} > \Lambda_j^{(\mathrm{Opt})}$, and the theorem follows.
\end{proof}

\begin{corollary}
$\Lambda_j^{(\mathrm{NEP})} {=} \Lambda_j^{(\mathrm{Opt})}
{\iff}
\Lambda_j^{(\mathrm{Opt})} {=} 0$.
\end{corollary}

\subsection{Price of anarchy}

The PoA in the studied system is the ratio between average latency at the NEP and average latency at the optimum. 
It is a function of the offered traffic---denoted as $\eta(\Lambda)$---and can only assume values greater than or equal to 1: 
\begin{align}
    \eta(\Lambda) 
    & =
    \frac
        { U ( \mathbf{p^{\dagger}} ) }
        { U ( \mathbf{p^{\star}} ) }
    =
    \frac
        { \alpha }          
        { U ( \mathbf{p^{\star}} ) }
    \ge 1.
\end{align}

\begin{theorem}
The PoA curve vs $\Lambda$ is piece-wise convex.
\label{th:poa+concave}
\label{th:concave}
\end{theorem}
\begin{proof}
Consider the adjacent segments $[\Lambda_j^{(\mathrm{NEP})},  \Lambda_{j+1}^{(\mathrm{NEP})}]$.
At $\Lambda_j^{(\mathrm{NEP})}$, server $j$ is activated and the NEP solution assigns probabilities $\mathbf{p^{\dagger}}$ so that the curve of $\alpha$ vs $\Lambda$ be continuous. However, the NEP does not assure continuity of the derivative of $\alpha$. Indeed, immediately before and after the activation of server $j$ the same latency is observed with a different number of servers, so that the last activated server $j$
absorbs some load and the rest of servers observe  a slowdown in the rate of increase of their latency. This corresponds to a sudden decrease of the derivative of $\alpha$. 
Instead, the condition used to compute $\gamma$ at the optimum preserves the continuity of the derivative of the latency. 
This means that while $\alpha$ vs $\Lambda$ experiences a drop in its growth rate at any point at which a new server is incorporated in the NEP, the latency at the optimum does not observe such a discontinuity in the derivative. The result is that, for arrival rates slightly larger than $\Lambda_j^{(\mathrm{NEP})}$, the PoA can decrease. However, the growth rate of $\alpha$ must quickly increase again and faster than the growth rate of the latency at the optimum because the NEP does not allow any server, not even the faster, to experience less latency than the slower. Therefore, the growth rate of $\alpha$ is that of the slowest active server, while at the optimum this cannot occur by construction. Since the numerator of the PoA increases faster than the denominator, the portion of the PoA curve between two consecutive server activation events at the NEP must be convex. This behavior holds for all feasible load segments, including from $\Lambda_n^{(\mathrm{NEP})}$ to $\sum_{j=1}^n \mu_j$. 
\end{proof}

\begin{corollary}
The maximum of the PoA vs $\Lambda$ occurs at a point of activation of a server at the NEP or at $\rho=1$.
\label{th:discretemax}
\end{corollary}

Corollary~\ref{th:discretemax} ensures that finding the max of the PoA can be done by evaluating a finite set of points. Hence, the cost of evaluating the worst case behavior of the distributed approach is comparable to the complexity of the exact algorithms for the computation of optimum and NEP. 

\section{Special cases}
\label{SPECIAL}

With \textbf{M/M/1 queues},
$\ell_i$ and $h_i$ can be inverted in closed form since they have simple expressions as follows:
\begin{align}
\ell_i (x) & = d_i + \frac{1}{\mu_i - x}; \qquad 
h_i(x) 
& = 
d_i + \frac{\mu_i}{ \left(\mu_i - x\right)^2}.
\label{eq:derivative}
\end{align}

\noindent General case expressions for optimization simplify into:   
\begin{align}
& \gamma = d_j + \frac{\mu_j}{\left(\mu_j - p_j^{\star} \Lambda\right)^2},\quad  p_j^{\star} >0, \quad \forall j \le j^{\star};
\\
& p_j^{\star} =  \frac{1}{\Lambda}\left(\mu_j - \sqrt{\frac{\mu_j}{\gamma-d_j}}\right), \quad \forall j \le j^{\star}; 
\label{eq:pj}
\\
&  \frac{1}{\Lambda} \sum_{j = 1}^{j^{\star}} 
    \left(\mu_j - \sqrt{\frac{\mu_j}{\gamma - d_j}}\right) = 1. 
    \label{eq:gamma_invert}
\end{align}
Notice that the complexity of inverting~\eqref{eq:gamma_invert} is 
$\mathcal{O}(\log r)$ instead of $\mathcal{O}(n\log r)$ observed in the general case, because the inversion can be done over the sum directly.
However, the overall complexity of the exact optimization remains $\mathcal{O}(n^2 + n\log r)$. 

\noindent
The expressions needed to study the NEP become: 
\begin{align}
    & \alpha = d_j + \frac{1}{\mu_j - p_j^{\dagger}\Lambda}, \quad  p_j^{\dagger} >0, \quad \forall j \le j^{\dagger};
\\
    & p_j^{\dagger} = \frac{1}{\Lambda} \left(\mu_j - \frac{1}{\alpha-d_j}\right), \quad \forall j \le j^{\dagger}; 
\\
    & \frac{1}{\Lambda} \sum_{j=1}^{j^{\dagger}} \left(\mu_j - \frac{1}{\alpha-d_j}\right) = 1.
\end{align}

\noindent As observed for the general case, 
the maximum value of the PoA can be found at any of the NEP activation points or at the limit for $\Lambda \rightarrow \sum_{j=1}^n \mu_j$. 
With M/M/1 queues, the latter can be computed in closed form.

\begin{theorem}
With M/M/1 queues, 
\begin{align}
    \lim_{\Lambda \rightarrow \sum_{j=1}^{n} \mu_j} \eta(\Lambda) 
    =
    n\, 
    {\sum_{j=1}^{n} \mu_j}
    \Bigg /{\left(\sum_{j=1}^{n} \sqrt{\mu_j}\right)^2}.
    \label{eq:poa_mm1_asymptotic}
\end{align}
\end{theorem}
\begin{proof}[Sketch of the proof]
The average latency at the NEP tends to diverge and can be computed in closed form by neglecting the fixed delay. 
By writing the expression of  $p_j^{\star}$ vs $\gamma$ in near-saturation conditions, with $\gamma {\gg} d_j$, and by summing all probabilities, we can derive the asymptotic expression for $\gamma$, hence derive $p_j^{\star}$ and the average latency at the optimum.
\end{proof}

In case of equal delays $d_j = d^{\star}$, closed forms can be found from the above formulas, 
which generalizes the conclusions of
Haviv and Roughgarden~\cite{HAVIV_ROUGH}, who studied the optimization and NEP of systems with M/M/1 queues with $d^{\star}=0$.

\vspace{2mm}
With \textbf{M/G/1 queues}, the following expressions hold as long as the queues are stable (i.e., for $x < \mu_j$): 
\begin{align}
\ell_j (x) & = d_j + \frac{1}{\mu_j} \left(1 + \frac{1 + C_j^2}{2} \frac{x}{\mu_j-x}\right)
,
\\
h_j(x) & =  
d_j + \frac{1}{\mu_j} 
\left(1 + \frac{1 + C_j^2}{2} \, \frac{2\mu_j-x}{\left(\mu_j-x\right)^2} \, x\right),
\end{align}
with
$C_j$ the coefficient of variation of server $j$'s service time.  

\noindent 
At the optimum and at the NEP we have:
\begin{align}
& p_j^{\star} 
{=}
\! \frac{\mu_j}{\Lambda} \left(\!1 {-} \frac{1}{\sqrt{1 {-} \frac{2}{1+C_j^2} \left(1{-}\mu_j \left(\gamma {-} d_j\right)\right)}}\right)\!\!, \quad \forall j {\le} j^{\star}\!;
\\
&    \frac{1}{\Lambda}
    \sum_{j=1}^{j^{\star}}
    \mu_j  
    \left(1 - \frac{1}{\sqrt{1- \frac{2}{1+C_j^2} \left(1-\mu_j \left(\gamma - d_j\right)\right)}}\right)
    = 1;
\\
& p_j^{\dagger} {=} \frac{\mu_j}{\Lambda}
    \!\left(\!
    1 - 
    \frac{1}{1-\frac{2}{1+C_j^2}\left(1-\mu_j(\alpha-d_j)\right)}
    \right)\!\!, \quad \forall j {\le} j^{\dagger};
\\
& \frac{1}{\Lambda}
\sum_{j=1}^{j^{\dagger}}
\mu_j
\left(
    1 - 
    \frac{1}{1-\frac{2}{1+C_j^2}\left(1-\mu_j(\alpha-d_j)\right)}
    \right) = 1. 
\end{align}
%



\noindent The asymptotic expressions for the PoA is as follows:
%
\begin{align}
    \eta & \simeq \frac{\Lambda \, \sum_{j=1}^n \frac{1+C_j^2}{2}}{\left(\sum_{j=1}^{n} \sqrt{\mu_j\,\frac{1+C_j^2}{2}}\right)^2}
    \rightarrow 
    \frac{  \sum_{j=1}^n \frac{1+C_j^2}{2} \, \sum_{j=1}^{n} \mu_j}
    {\left(\sum_{j=1}^{n} \sqrt{\mu_j\,\frac{1+C_j^2}{2}}\right)^2}. 
    \label{eq:eta_mg1}
\end{align}
It is interesting to notice that the asymptotic behavior of the PoA differs from the M/M/1 case due to terms $(1+C_j^2)/2$. Moreover, as it can be expected, setting $C_j=1$ for all servers yields the results previously found for the M/M/1 queue, while  $C_j=0$ would lead to results valid for the M/D/1 queue.

\section{Performance evaluation}
\label{RESULTS}
\subsection{Experimental apparatus}

We set up a distributed measurement apparatus with the aim of obtaining a dependable ground truth to contrast with the predictions of our model. 
The apparatus enables the configuration of a group of servers (in our experiment, three servers were utilized) with varying service capacities and diverse network locations. 
This tool is coded in Golang \cite{golang}, utilizing microservices for deployment, and employs QUIC as transport layer protocol \cite{quic}.
The tool specifies three primary entities: clients, servers, and routing nodes. The server instantiates a configurable web server that exposes one or multiple services with distinct computational demands. Client and server support a fine-tuning of applications and traffic shape characteristics. The routing element manages the traffic generated by other entities, directing it to the designated destination.

The tool collects and stores networking and traffic routing events, including packet arrivals and departures to and from various elements, real-time monitoring of memory occupancy status, and the dynamic count of available threads dedicated to handling incoming traffic at the servers. It also allows for easy deployment of application-specific measurements.

The experimental setting comprises three servers deployed across Europe, each characterized by deterministic service times, practically no bounds on available memory, and one working thread. 
The application requires clients to issue some small update packets (100 bytes) to any of the available servers; the server computes the updated information or instruction for each client and then sends it back to her. 

In our setting, several co-located client elements issue packets to the three servers; the first two elements generate a traffic load equal to 5\% of the overall system capacity---and this is the traffic being measured---while the last element provides background traffic. 
In the experimental evaluation, we measure the round trip time between each client and the three servers; then, we use it to compute the allocation of traffic to servers, and we feed this configuration to the routing element and run the experiments.
In the case of the game, this is equivalent to saying that rational players can do the math and choose their actions at the NEP without actually playing.


\subsection{Results and validation}

We used a few significant scenarios to validate our analytical model and showcase the analytical and algorithmic results derived in this paper.

\begin{figure}[!t]
\centering
\includegraphics[width=0.9\columnwidth]{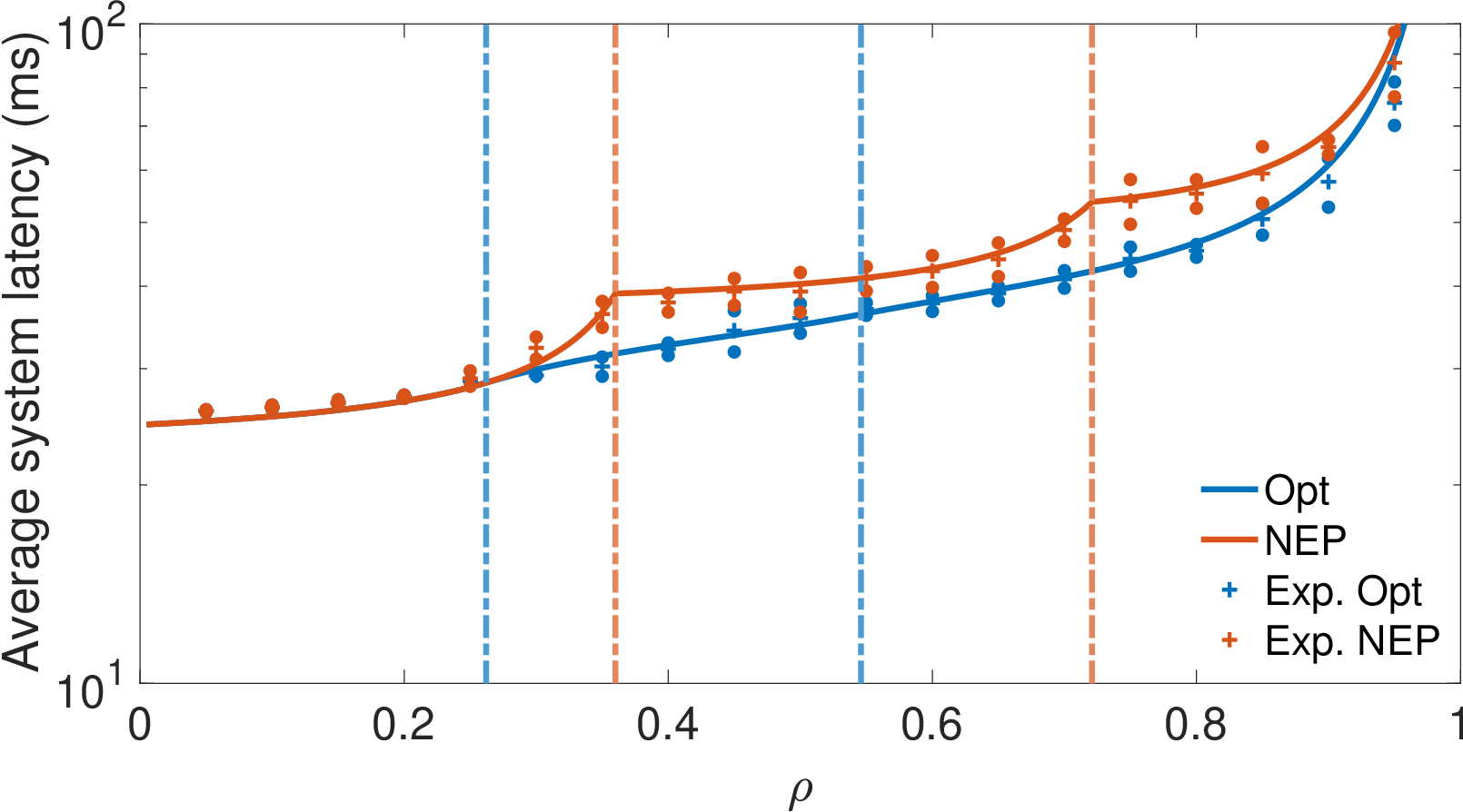}
\vspace{-3mm}
\caption{Validation in Scenario 0}
\vspace{2mm}
\label{fig:experimental_md1}
\end{figure}

\begin{figure}[!t]
\centering
\includegraphics[width=0.9\columnwidth]{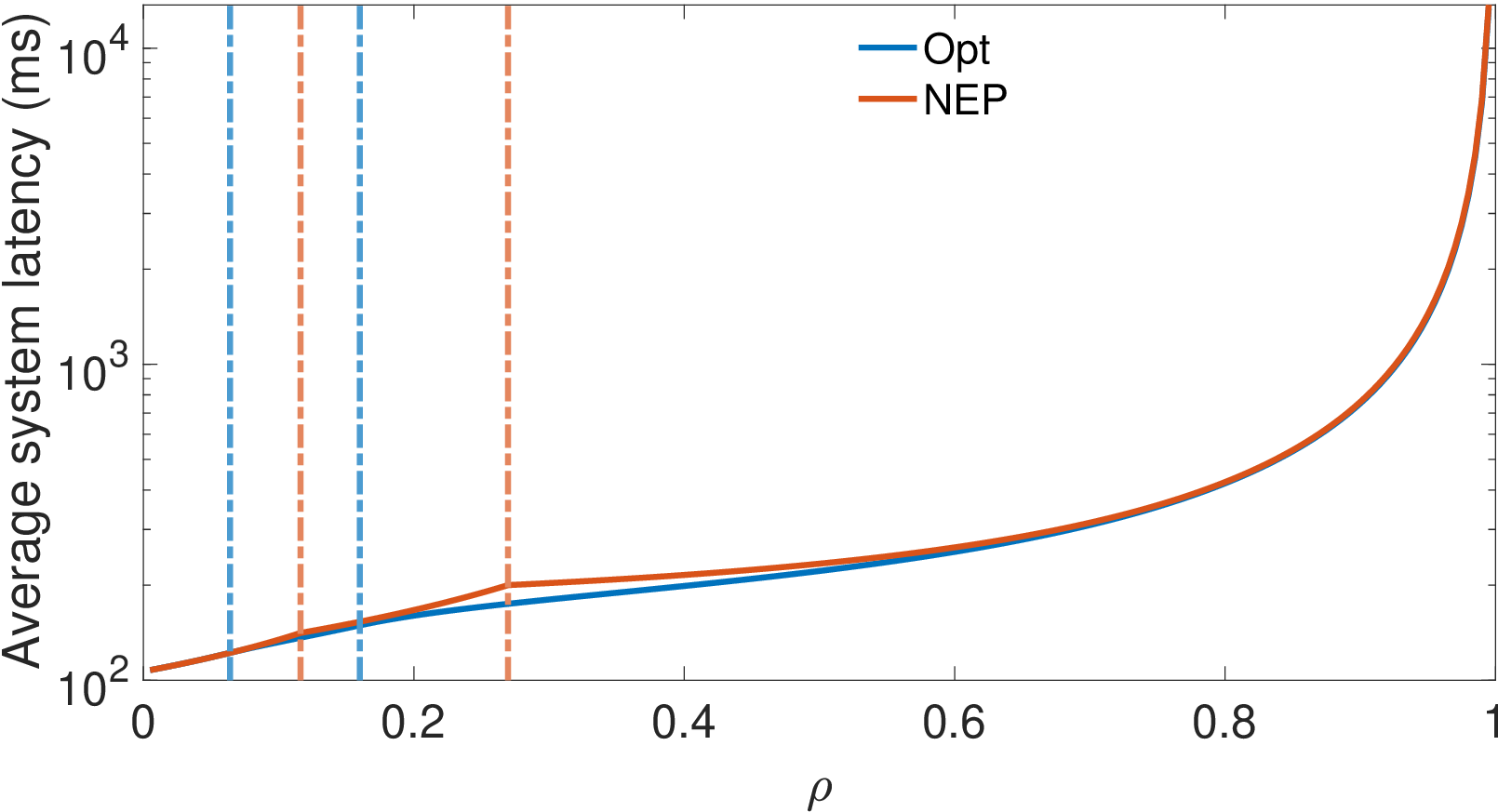}
\vspace{-3mm}
\caption{Latency vs offered load in Scenario 1}
\vspace{2mm}
\label{fig:latency_s1}
\end{figure}

\begin{figure}[!t]
\centering
\includegraphics[width=0.9\columnwidth]{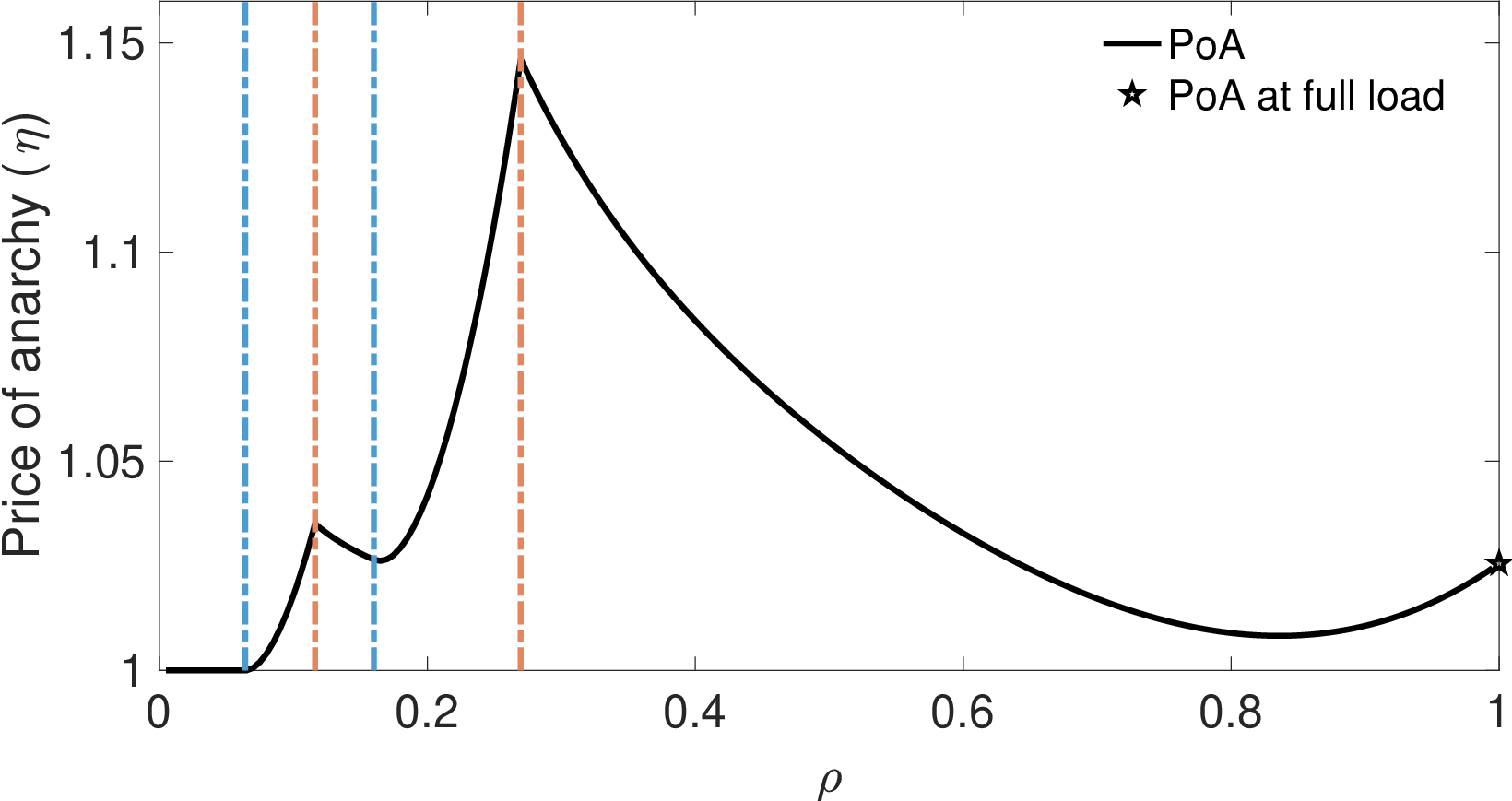}
\vspace{-3mm}
\caption{Price of anarchy vs load in Scenario 1. The marked point indicates the value computed with~\eqref{eq:poa_mm1_asymptotic}.}
\label{fig:poa_s1}
\end{figure}

\textbf{Scenario 0---Validation.}
We start by validating our model via experiments in a real network context.
We observed the two-way delay between users and servers with ping packets, and we used the average RTT (round trip time) values to run our analytical model. The characteristics of the available
servers, in order of activation, are $d = [20,\, 34,\, 43.5]$ ms, and $ \mu = [4.66,\, 5.00,\, 10.20
]$ services/s. 

Fig.~\ref{fig:experimental_md1} reports the results generated by the model and the experiments (with 90\% confidence intervals).
The figure also shows the activation thresholds of the servers, with vertical dotted lines (using the same color as the corresponding latency curves). Thresholds are computed analytically, based on the observed average two-way delay.  

Analytical results show that performance differences between the optimum and the NEP are generally not large and can only be experienced when at least two servers are active, because with one active server there is nothing to optimize.

Experimental results are well approximated by the analysis, which tells that considering a constant two-way delay instead of using a stochastic model yields an affordable simplification.

\textbf{Scenario 1---M/M/1 Edge \& Cloud.}
In this scenario we consider three M/M/1 servers with different capacity. One of the three servers is much farther apart from users than the other two servers, but it is faster. This scenario represents a case in which two servers are within the edge area of the network and one is in the cloud. The characteristics of the available servers, in order of activation, are $d=[40,\,30,\,150]$ ms, and  
$\mu=[15,\,9,\,20]$ services/s. 
Latency and PoA as a function of the offered load in this scenario are shown in Fig.~\ref{fig:latency_s1}
and Fig.~\ref{fig:poa_s1}, respectively. 
 As the second server gets activated at the optimum, the latency at the NEP starts increasing faster than in the optimized system, and the PoA becomes larger than 1. However, the activation of the second server at the NEP causes a temporary decrease in the PoA, after which it goes up quickly. Indeed, the PoA curve is visibly piece-wise convex and the peak of the PoA is reached at the activation of the last server at the Nash equilibrium. The corresponding value is neither negligible nor very large (below 1.15). It is interesting to notice that the PoA can be quite different at different loads, and that a distributed implementation of the job allocation could work well at quite high loads, in this scenario. However, this result does not apply in general, as we will see with the analysis of the following scenario.

\begin{figure}[!t]
\centering
\includegraphics[width=0.9\columnwidth]{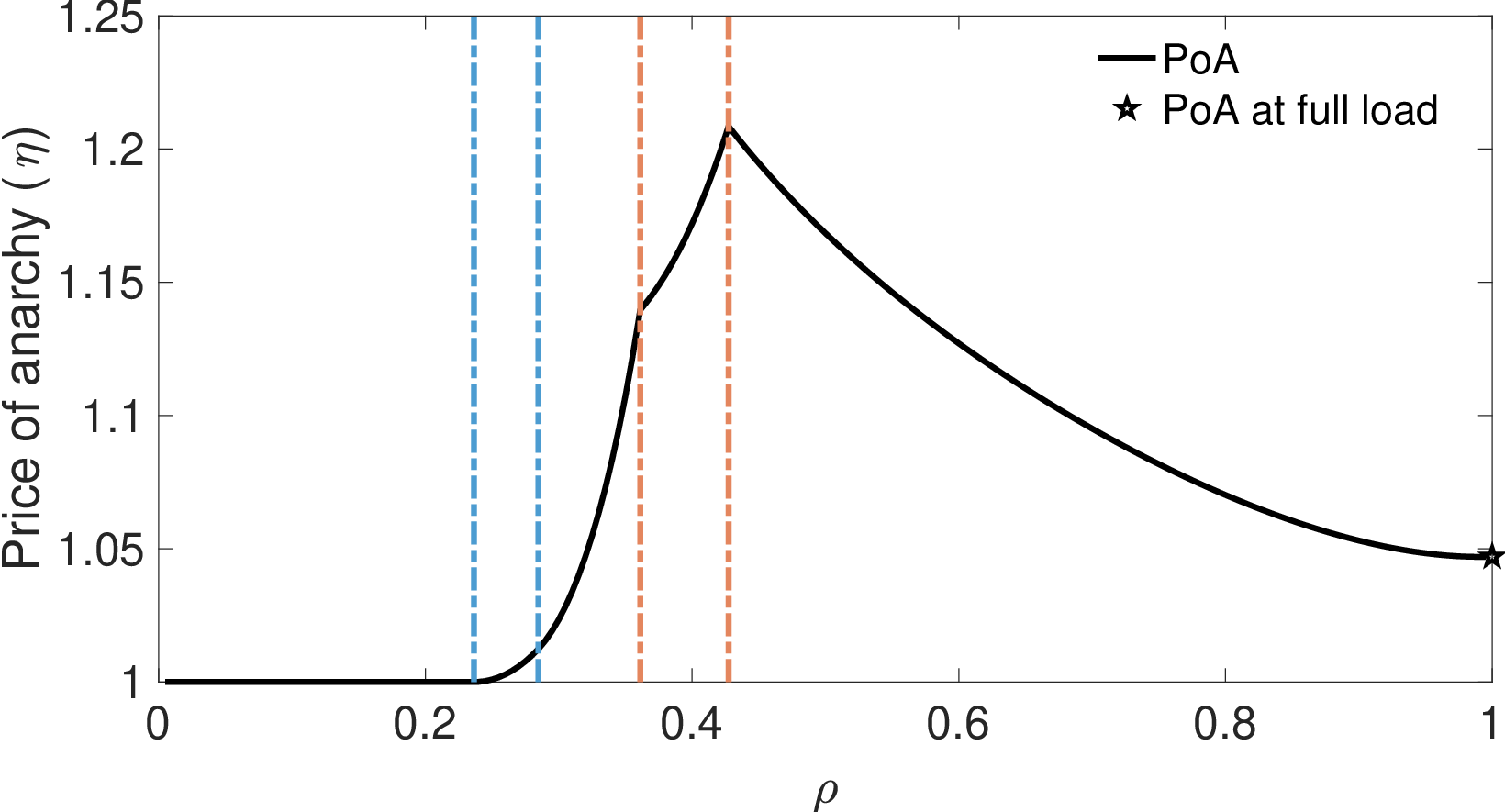}
\vspace{-3mm}
\caption{Price of anarchy vs load in Scenario 2. The marked point indicates the value computed with~\eqref{eq:poa_mm1_asymptotic}.}
\label{fig:poa_s2}
\end{figure}

\textbf{Scenario 2---M/M/1 Heterogeneous Edge.}
We use three M/M/1 edge servers, all of them close to users, with important differences in the  capacity. In order of earlier server activation, we have 
$d=[10,\,12,\,20]$ ms, and  
$\mu=[300,\,100,\,200]$ services/s. 
Results for this scenario are shown in 
Fig.~\ref{fig:poa_s2}. 
In this case, the optimization activates all servers before that the NEP dictates the activation of the second server. This happens because the second server saturates quickly.

It is interesting to notice that in this case the PoA increases also after the activation of the second server at the NEP, which happens because the second server is activated at the NEP after the optimum has activated the third and last server, hence, the capacity used at the optimum remains higher than the one used at the NEP until all servers are active also for the NEP.

In this case, the shape of the PoA curve is very different with respect to Scenario 1, and a distributed implementation might be suitable only for  low loads. 
In general, the shape of $\eta$ vs $\rho$ is a concatenation of convex functions, although predicting the position and value of the peak is not straightforward, as the shape of the curve can change significantly with the distance and capacity of the available servers.

\begin{figure}[!t]
\centering
\includegraphics[width=0.9\columnwidth]{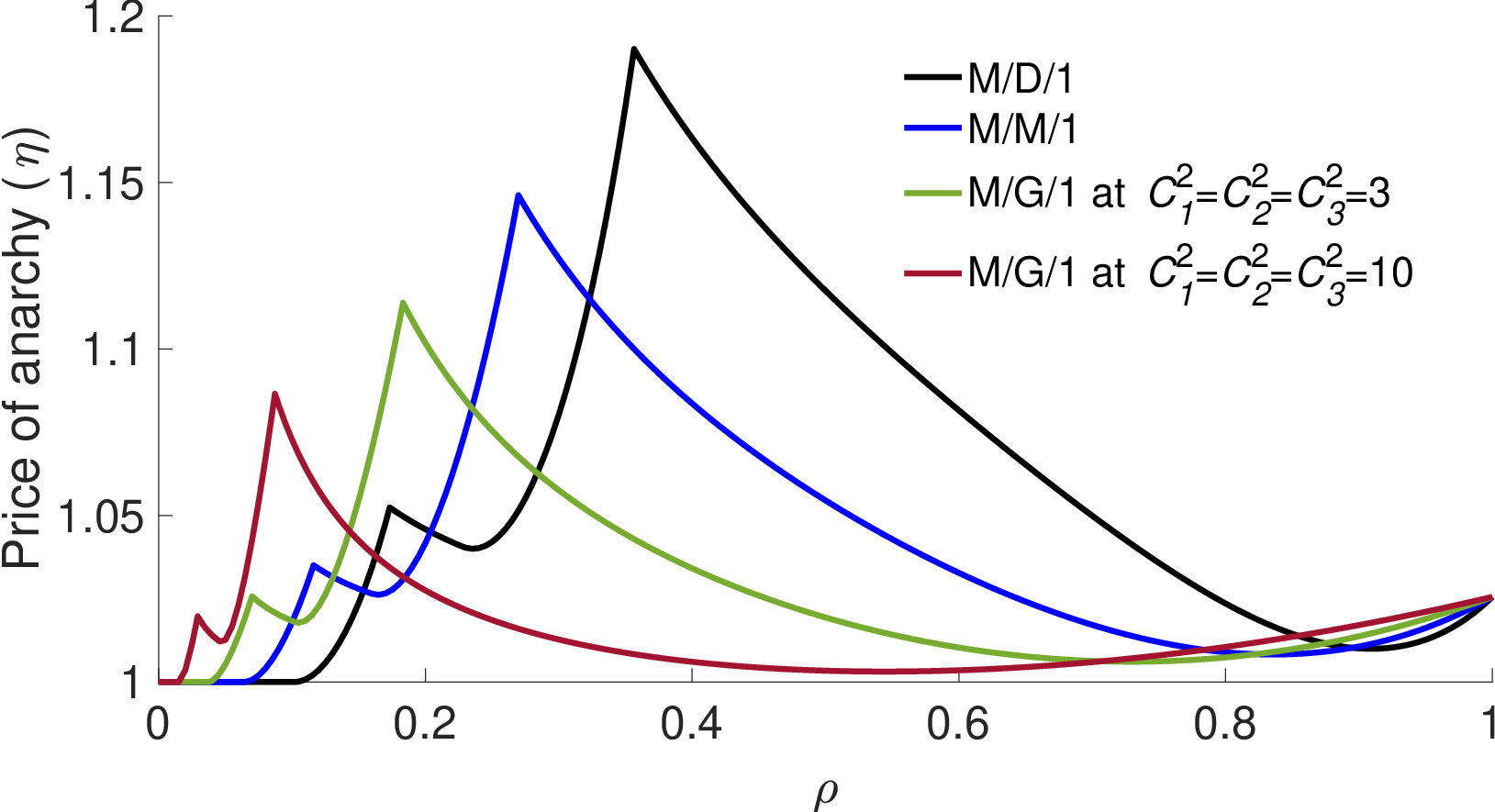}
\vspace{-3mm}
\caption{Scenario 3: Price of anarchy vs load at various levels of variance (the coefficients of variations are the same for all servers)}
\vspace{2mm}
\label{fig:poa_s3_multi}
\end{figure}

\begin{figure}
\centering
\includegraphics[width=0.9\columnwidth]{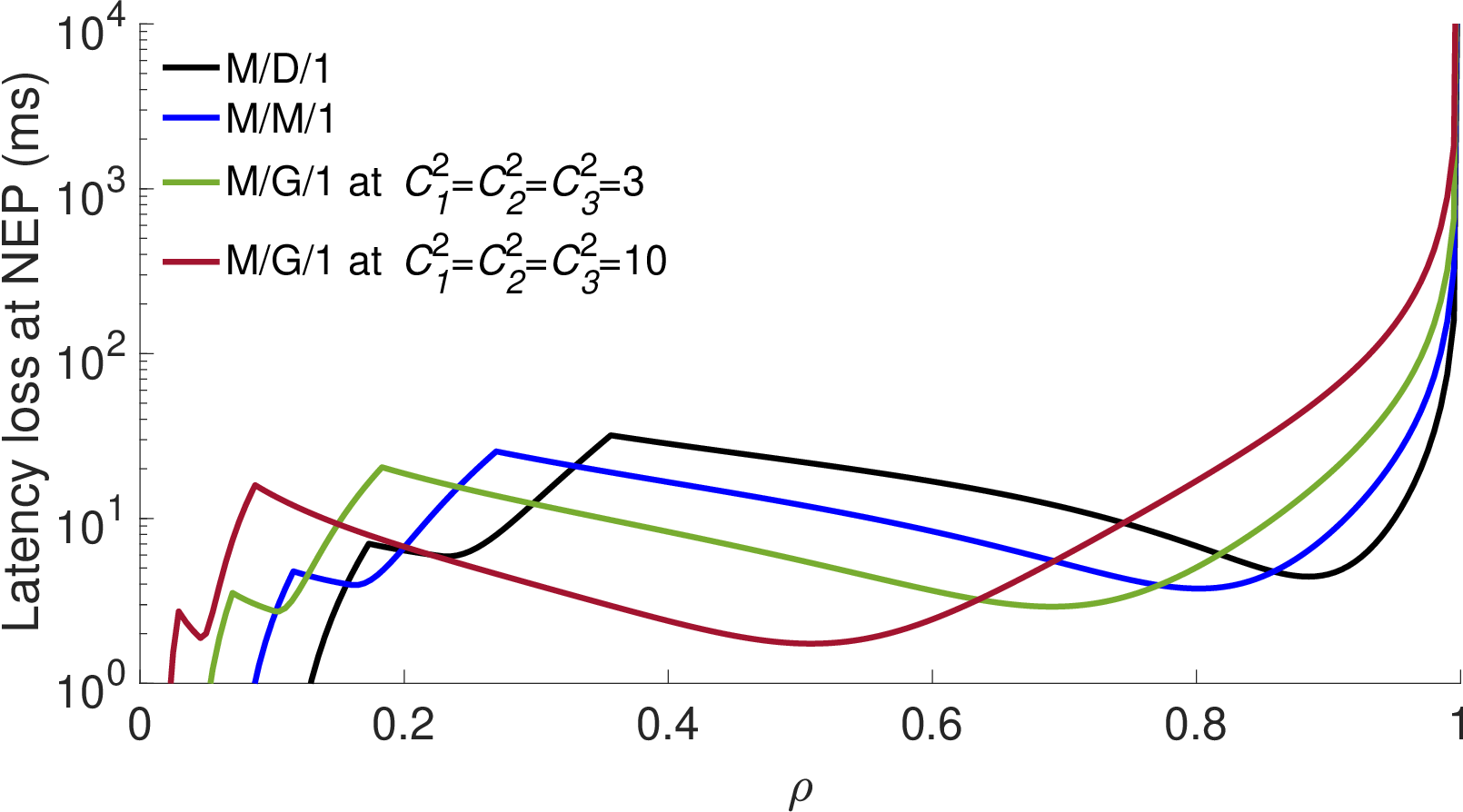}
\vspace{-3mm}
\caption{Scenario 3: Additional delay incurred by NEP, with respect to the optimum, vs load at various levels of variance (the coefficients of variations are the same for all servers)}
\label{fig:poadiff_s3_multi}
\vspace{2mm}
\end{figure}

\begin{figure}[!t]
\centering
\includegraphics[width=0.9\columnwidth]{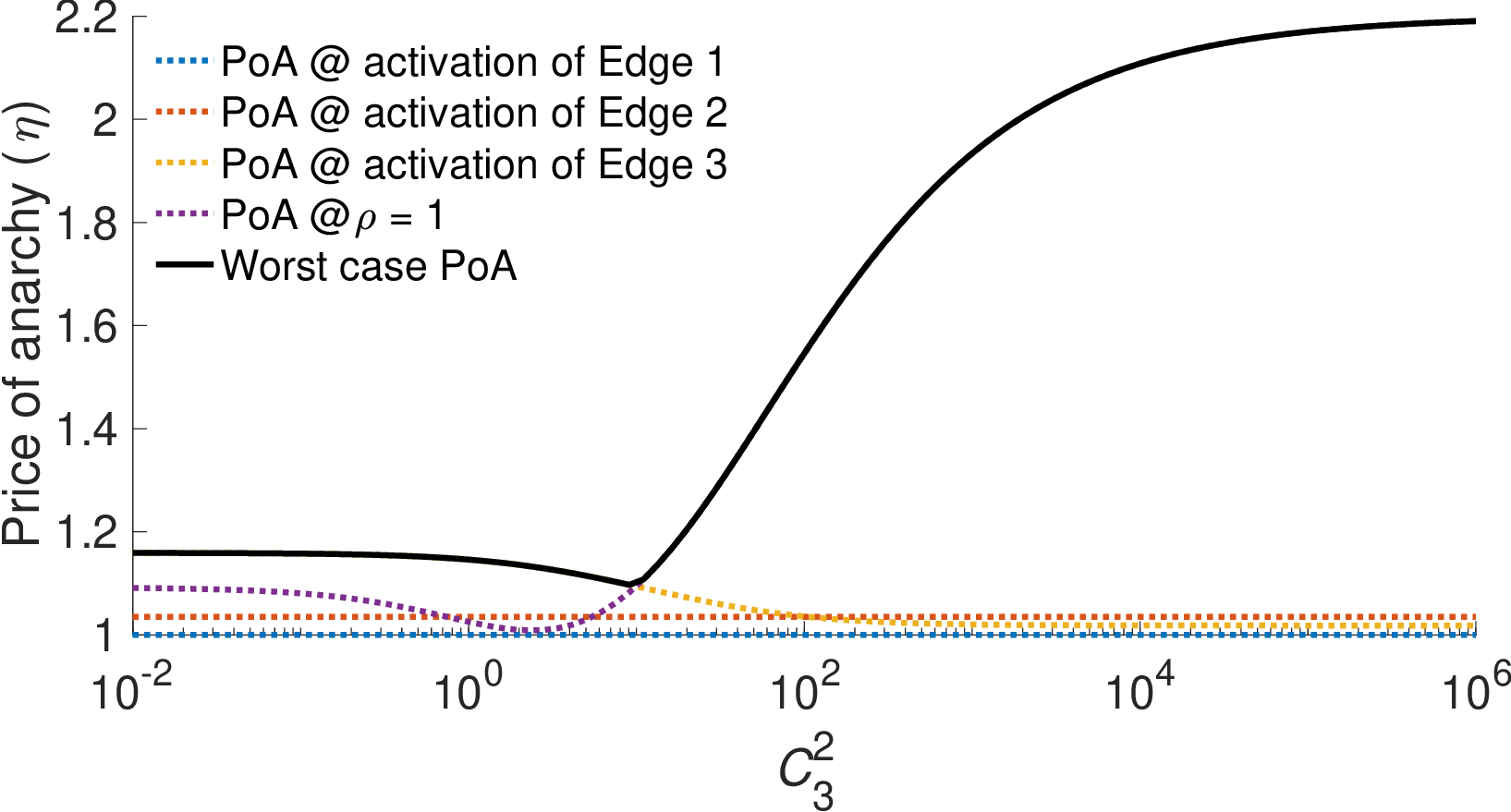}
\vspace{-3mm}
\caption{Highest price of anarchy vs the coefficient of variations of the server farther apart (Edge 3)}
\vspace{-2mm}
\label{fig:poa_vs_C_far}
\end{figure}

\textbf{Scenario 3---Impact of variance.}
We consider servers as in Scenario 1, except we now use M/G/1 queues and compare the impact of the coefficients of variation. Results are reported in Figs.~\ref{fig:poa_s3_multi}
and 
\ref{fig:poadiff_s3_multi}. 
The figures include the special cases of M/D/1 and M/M/1 servers, obtained by using coefficients of variation equal to 0 and 1, respectively. They also include two cases with high and very high variance, obtained by setting all coefficients of variation to the same value, which is either 3 or 10.

We can see that the peak of the PoA decreases with the variance, as shown in  Fig.~\ref{fig:poa_s3_multi}. It is also interesting that the PoA peak moves to the right of the graphs (i.e., it occurs at higher loads) as the variance of the service time decreases. This happens because the denominator of the PoA increases with the variance, hence partially hiding the excess latency due to operating the service at the NEP.
Indeed, Fig.~\ref{fig:poadiff_s3_multi} shows that the net latency loss due to a distributed implementation, at the NEP, has peaks that are comparable for the different cases, and tends to be very large only when the load approaches 1.

Eventually, with the experiment of Fig.~\ref{fig:poa_vs_C_far}, we discuss the impact of the variance observed at a single server while the rest of servers use M/M/1 queues. Dotted lines in the figure show the value of the PoA points that are candidate for the worst case, i.e., NEP activation points and full load. As shown in the figure, the worst case (solid line) is dominated by one of the activation points until the point at full load takes the lead and the system starts behaving far from optimally. Here we have reported the case in which we vary the coefficient of variation of the server farther apart, but the behavior is qualitatively the same for all servers. The curve of the worst case PoA has an asymptotic value which is readily obtained from \eqref{eq:eta_mg1}, whose value is $\sum_k \mu_k / \mu_j$, for the case in which we vary $C_j^2$. However, the fact that the worst PoA can be high when it is dominated by the behavior at full load is not critical for two reasons: first, the system should not run close to saturation and, second, other candidate points for the worst PoA value never yield high PoA values, which means that a service that runs at reasonable load will never suffer high PoA.

\begin{figure}[!t]
\centering
\includegraphics[width=0.9\columnwidth]{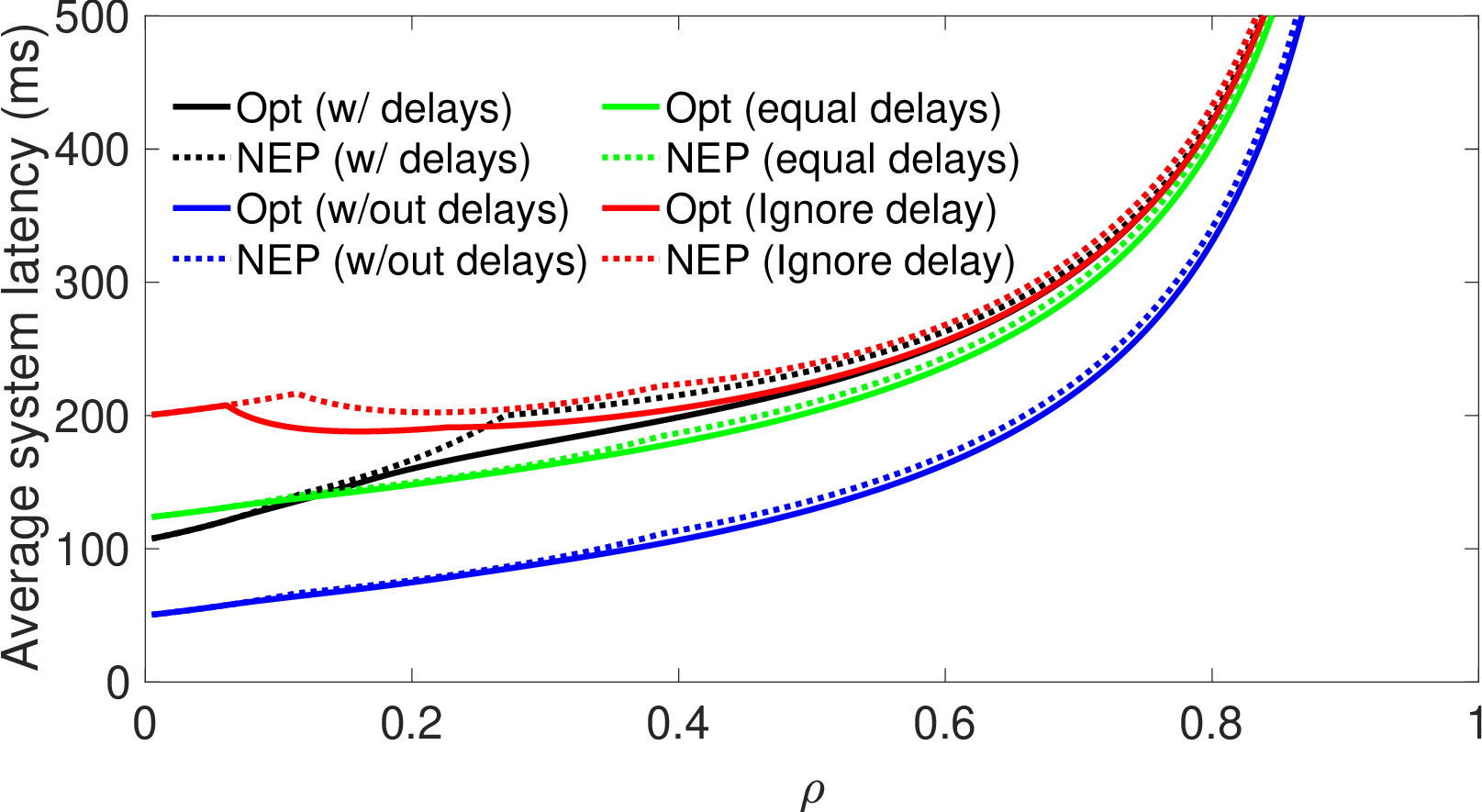}
\vspace{-3mm}
\caption{Latency vs offered load in Scenario 1 with variants on the evaluation and consideration of fixed delays}
\vspace{2mm}
\label{fig:latency_compare_mm1}
\end{figure}

\begin{figure}
\centering 
\includegraphics[width=0.9\columnwidth]{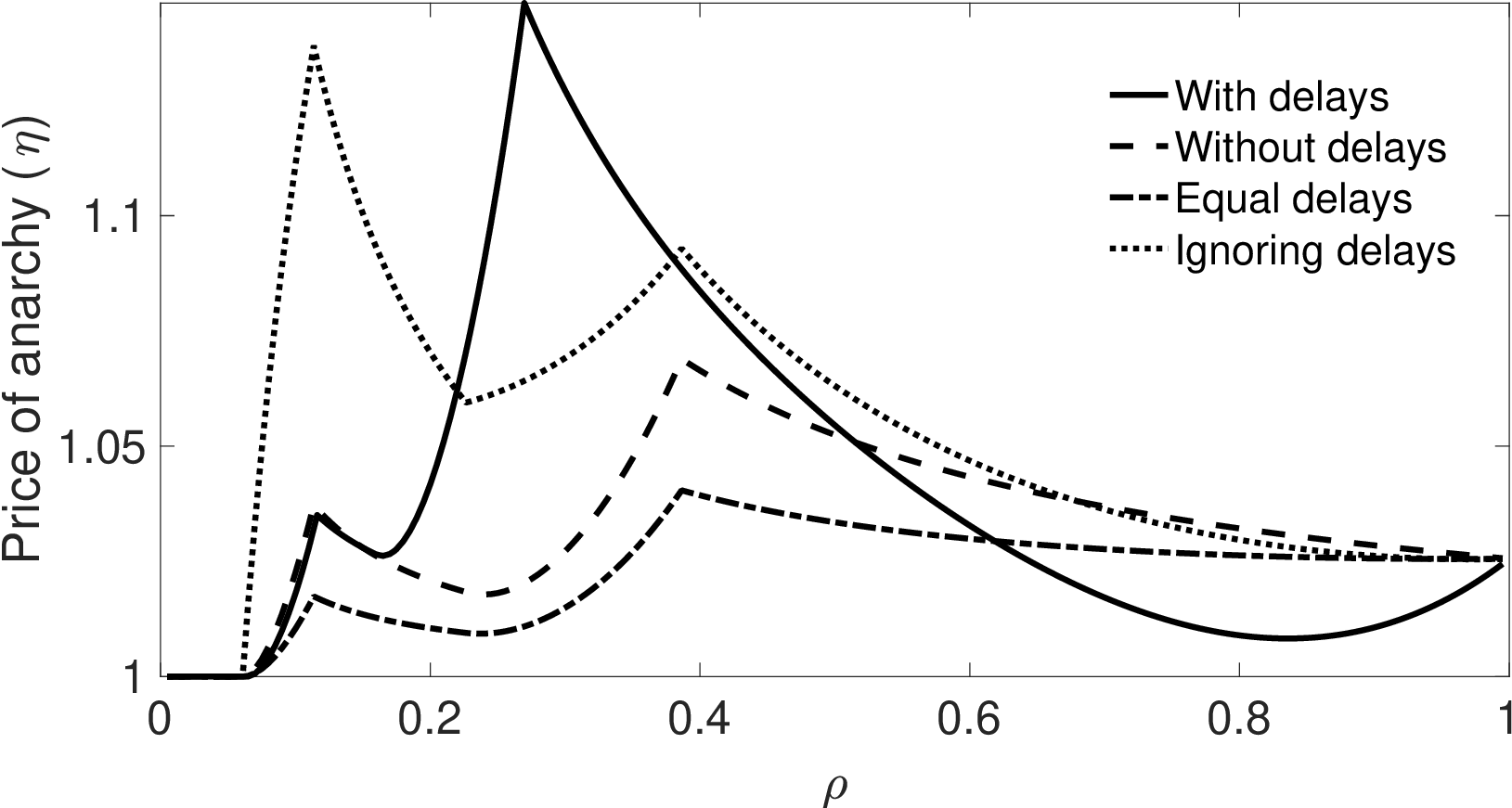}
\vspace{-3mm}
\caption{Price of anarchy vs load in Scenario 1 with variants on the evaluation and consideration of fixed delays}
\label{fig:poa_compare_mm1}
\end{figure}

\textbf{Scenario 4---The importance of accounting for fixed delays.} 
Fig.~\ref{fig:latency_compare_mm1}
and Fig.~\ref{fig:poa_compare_mm1} report results for a network configuration like the one of Scenario 1, although the optimization and the NEP are computed under different assumptions. Specifically, we consider a first variant in which we  neglect the existence of fixed delays in the optimization algorithm and in the NEP calculation (assuming that the user is not actually observing her latency but simply trusting her math). This case is labeled as ``Ignoring delays'' in the figures, and shows large errors with respect to the correct optimization or NEP calculation. Thus, accounting for fixed delays makes a huge difference. 
We further compare the results discussed for Scenario 1 with the case in which fixed delays $d_{j}$ are set to zero (``Without delays'', in the figures). In this case, results are sensibly better than in the originally considered setup for Scenario 1 (``With delays'', in the figures), at least in terms of latency, and only until the load becomes high. This means that, again, fixed delays play an important role and neglecting them is not possible before the average sojourn time into servers becomes predominant.    
Another variant is obtained by considering a modified Scenario 1 with fixed delays which are the same for all servers. We consider the average value of the original Scenario 1 (i.e., $d_{j}{=}73.3$ ms for all servers). Latency curves in this case are smoother, and the difference with Scenario 1 vanishes as the load approaches 1.  

Note that the dotted curve in the figure, obtained by ignoring the fixed delays in optimization and NEP calculation is not even convex. However,  delays are present, so they are properly accounted for in the aftermath of the optimization/NEP evaluation.  
This result does not contradict Theorem~\ref{th:concave} as the dotted curve is the result of an inaccurate mathematical approximation that could lead to inaccurate optimizations and/or NEP calculations---for the latter case, selfish users acting distributedly can easily see it.
The dotted curve is intentionally included with the purpose to show that inaccurate assumptions might not only lead to wrong decisions, but also hide important properties.  

In all cases, the differences in terms of the maximum value observed in the PoA are limited, although $(i)$ when delays are uniform, the PoA tends to be smaller and $(ii)$ the shapes of the curves are very different, with peaks occurring at very different values of the load.

\section{Conclusions}
\label{CONCS}
The analysis and optimization of computing task allocation in the edge-cloud continuum requires attention to latency characteristics whose effect was traditionally neglected. We have shown that the analysis of the latency experienced by end-users becomes more complex than in previously studied systems, yet its optimization and NE can be analytically characterized and computed with exact algorithms that we have derived in this paper. The analysis we presented is very general, and only requires that the sojourn time of a task in a server be a convex function of the load of the server, which is a common property of all non-fully-deterministic systems.  

Our findings were validated through the deployment of a real distributed system which spans a multiparty laboratory across different countries. Our results have shown that $(i)$ optimal configurations and selfish equilibria are not necessarily intuitive to find, as they exhibit a strong dependency on relative differences between available servers, in terms of capacity and distance from the user, and $(ii)$ distributed and selfish optimizations incur limited performance costs, unless the system is driven into deep saturation and the variance of the service time becomes unrealistically high. 

\bibliographystyle{IEEEtran}
\bibliography{IEEEabrv,PoA_arxiv}
\end{document}